\documentclass[aps,showpacs,prd,twocolumn]{revtex4}
\usepackage{amssymb}
\usepackage{amsfonts}
\usepackage{amsmath}
\usepackage{graphicx}
\usepackage{color}
\usepackage[dvips]{epsfig}
\usepackage[dvips]{graphicx}
\usepackage{float}

\begin{document}

\title{Topological charged BPS vortices in Lorentz-violating Maxwell-Higgs
electrodynamics}
\author{R. Casana, G. Lazar.}
\affiliation{$^{1}$Departamento de F\'{\i}sica, Universidade Federal do Maranh\~{a}o,
65080-805, S\~{a}o Lu\'{\i}s, Maranh\~{a}o, Brazil.}

\begin{abstract}
We have performed a complete study of BPS vortex solutions in the Abelian
sector of the standard model extension (SME). Specifically, we have coupled
the SME electromagnetism with a Higgs field which is supplemented with a
Lorentz-violating CPT-even term. We have verified that Lorentz violation
(LV) belonging to the Higgs sector allows to interpolate between some well
known models like, Maxwell-Higgs, Chern-Simons-Higgs and
Maxwell-Chern-Simons-Higgs. We can also observed that the electrical charged
density distribution is nonnull in both CPT-even and CPT-odd models;
however, the total electric charge in the CPT-even case is null, whereas in
the CPT-odd one it is proportional to the quantized magnetic flux. The
following general results can be established in relation to the LV
introduced in the Higgs sector: it changes the vortex ansatz and the gauge
field boundary conditions. A direct consequence is that the magnetic flux,
besides being proportional to the winding number, also depends explicitly on
the Lorentz-violation belonging to the Higgs sector.
\end{abstract}

\maketitle

\section{Introduction}

Physics in the context of Lorentz-violating theories have been an important
branch for different theoretical and experimental research in recent years.
The standard model extension (SME) \cite{Colladay1,Colladay2,Coleman1} is
the general theoretical framework for studying Lorentz violation (LV)
effects, and it is build by adding Lorentz-violating terms in all sectors of
the minimal standard model. Particulary, the Abelian gauge sector of the SME
is the principal focus in searching for LV effects in physical systems \cite%
{Jackiw}-\cite{Klink3}.

On the other hand, magnetic flux vortices have gained great interest since
Abrikosov's description for Type-II superconductors \cite{Abrikosov}, which
arise naturally from the non-relativistic limit of Ginzburg-Landau (GL)
theory \cite{GL}. In field theory, stable vortex configurations came up with
the seminal work by Nielsen and Olesen \cite{NO} whose study of the
Maxwell-Higgs (MH) model shows that electrically neutral vortex solutions
correspond to the ones obtained by Abrikosov. Lately, the existence of
electrically charged vortex solutions in Chern-Simons-Higgs (CSH) \cite%
{CS,CSV} and Maxwell-Chern-Simons-Higgs (MCSH) \cite{MCS,Bolog} models was
verified.

The study of topological defects in the context of Lorentz-violating models
has been performed in some cases,for example, scalar systems \cite{Defects},
Abelian monopoles \cite{Monopole1}, tensor fields \cite{Seifert}, etc. The
research for explicit BPS vortex solutions in the context of SME was
recently performed in Refs. \cite{Carlison1,Carlison2,Lucas,cpt13,lucas2}.

The aim of this article is to investigate how LV introduced in Higgs sector
affects the structure of topological defects; particularly, we discuss
charged BPS vortices in Lorentz-violating MH models. It is easily observed
that Higgs's Lorentz-violating term introduces anisotropies in the
dispersion relations of the Higgs field. Consequently, these anisotropies
provide a Hamiltonian dynamics very similar to GL models describing
anisotropic and layered superconductors (e.g. high-$T_{c}$ superconductors)
\cite{ref_1,ref_2,ref_3,ref_4}. So, the main motivation for studying
topological defects in the context of Lorentz-violating theories comes from
the similitude with the mathematical structure describing the physics of new
high-$T_{c}$ superconductors. These new structures lead to a strong
anisotropy of magnetic properties and gives rise to new phenomena, such as
formation of two-dimensional vortices and existence of new type of vortex
lattices \cite{ref_2,ref_4}.

The manuscript begins by studying the MH model supplemented with a CPT-odd
or Carroll-Field-Jackiw (CFJ) term and then with a CPT-even term in gauge
sector. In both situations we have introduced a CPT-even and
Lorentz-violating term in the Higgs sector. As can be expected, for the
first model electrically charged configurations appear due to the CFJ term
in the same manner as in MCSH model. For the second model, electric and
magnetic sectors are coupled due to parity-odd coefficients presents in the
electromagnetic CPT-even and Lorentz-violating term. In the following
sections, we analyze in detail both LV scenarios for
Abrikosov-Nielsen-Olesen (ANO) vortices.

\section{TOPOLOGICAL CHARGED BPS VORTICES IN A CPT-ODD AND LORENTZ-VIOLATING
FRAMEWORK}

We consider the Abelian sector of the SME, which corresponds to the MH model
supplemented with Lorentz-violating terms \cite{Colladay2}. Our first
analysis is restricted to the photon sector, including only the CFJ term
\cite{Jackiw}, while the Higgs sector is modified by a CPT-even term,
\begin{eqnarray}
\mathcal{L} &=&-\frac{1}{4}F_{\alpha \beta }F^{\alpha \beta }-\frac{1}{4}%
\epsilon ^{\alpha \beta \rho \sigma }\left( k_{AF}\right) _{\alpha }A_{\beta
}F_{\rho \sigma }  \label{Lag_CFJ} \\
&&+\left\vert D_{\mu }\phi \right\vert ^{2}+\left( k_{\phi \phi }\right)
^{\mu \nu }\left( D_{\mu }\phi \right) ^{\ast }\left( D_{\nu }\phi \right)
-U\left( \left\vert \phi \right\vert \right) ,  \notag
\end{eqnarray}%
where $A_{\mu }$ is the Abelian gauge field, and $F_{\mu \nu }=\partial
_{\mu }A_{\nu }-\partial _{\nu }A_{\mu }$ is the electromagnetic strength
tensor. The four-vector $\left( k_{AF}\right) _{\alpha }$ is the CFJ
background with mass dimension +1. The coupling between the gauge field and
Higgs field $\phi $ is given by the minimal covariant derivative $D_{\mu
}\phi =\partial _{\mu }\phi -ieA_{\mu }\phi $, where $e$ is the electric
charge. The LV in the Higgs sector is ruled by the CPT-even real symmetric
and dimensionless tensor $\left( k_{\phi \phi }\right) ^{\mu \nu }$. The
stationary Gauss law is%
\begin{equation}
\ \partial _{j}\partial _{j}A_{0}-\left( k_{AF}\right) _{i}B_{i}=-e\mathcal{J%
}_{0},  \label{gauss}
\end{equation}%
with $\mathcal{J}_{0}=\left[ 1+\left( k_{\phi \phi }\right) _{00}\right]
J_{0}-\left( k_{\phi \phi }\right) _{0i}J_{i}$, and $J_{\mu }=i\left[ \phi
\left( D_{\mu }\phi \right) ^{\ast }-\phi ^{\ast }D_{\mu }\phi \right] $ is
the conserved current density in absence of LV. Equation (\ref{gauss}) shows
clearly that the vectorial back-ground $\left( k_{AF}\right) _{i}$ is
responsible for the coupling between electric and magnetic sectors, allowing
the occurrence of electrically charged vortex solutions.

Similarly that happens in the MCSH model \cite{MCS,Bolog}, we must modify
the model in (\ref{Lag_CFJ}) by introducing a neutral scalar field $\Psi $
such that the new model supporting vortex solutions is
\begin{eqnarray}
\mathcal{L} &=&\mathcal{-}\frac{1}{4}F_{\alpha \beta }F^{\alpha \beta }-%
\frac{1}{4}\epsilon ^{\alpha \beta \rho \sigma }\left( k_{AF}\right)
_{\alpha }A_{\beta }F_{\rho \sigma }+\left\vert D_{\mu }\phi \right\vert ^{2}
\notag \\
&&+\left( k_{\phi \phi }\right) ^{\mu \nu }\left( D_{\mu }\phi \right)
^{\ast }\left( D_{\nu }\phi \right) +\frac{1}{2}\partial _{\mu }\Psi
\partial ^{\mu }\Psi  \\
&&-e^{2}\left[ 1+\left( k_{\phi \phi }\right) _{00}\right] \Psi
^{2}\left\vert \phi \right\vert ^{2}-U\left( \left\vert \phi \right\vert
,\Psi \right) ,  \notag
\end{eqnarray}%
and the potential $U\left( \left\vert \phi \right\vert ,\Psi \right) $
providing charged BPS vortices is
\begin{equation}
U=\frac{1}{2}\left[ ev^{2}-e\eta \left\vert \phi \right\vert ^{2}-\left(
k_{AF}\right) _{3}\Psi \right] ^{2},
\end{equation}%
where $\eta $ is defined in terms of the parity-even components of the
Lorentz-violating Higgs tensor
\begin{equation}
\eta =\sqrt{\left[ 1-\left( k_{\phi \phi }\right) _{\theta \theta }\right] %
\left[ 1-\left( k_{\phi \phi }\right) _{rr}\right] }.
\end{equation}%
It can be observed that this potential provides two vacua states, one
symmetric $\left[ \phi =0,\,~\Psi =ev^{2}/\left( k_{AF}\right) _{3}\right] $
and another asymmetric, responsible for generating topological BPS vortices\
$\left[ \Psi =0\mathbf{,\ }\left\vert \phi \right\vert =v/\eta ^{1/2}\right]
$. The Higgs field vacuum expectation value gains contributions from
Lorentz-violating terms introduced in Higgs sector.

\subsection{BPS formalism}

We are interested to find self-dual equations or BPS \cite{BPS} which are
first-order differential equations whose solutions also solve the
second-order Euler-Lagrange equations guaranteeing minimum energy
configurations. To carry out the BPS procedure, we first calculate the stationary
energy density of the model,
\begin{eqnarray}
\mathcal{E} &=&\frac{1}{2}B^{2}+\frac{1}{2}\left[ ev^{2}-e\eta \left\vert
\phi \right\vert ^{2}-\left( k_{AF}\right) _{3}\Psi \right] ^{2}  \notag \\%
[0.2cm]
&&+\left[ \delta _{jk}-\left( k_{\phi \phi }\right) _{jk}\right] \left(
D_{j}\phi \right) ^{\ast }\left( D_{k}\phi \right) +\frac{1}{2}\left(
\partial _{k}A_{0}\right) ^{2}  \label{ED1} \\[0.2cm]
&&+\frac{1}{2}\left( \partial _{k}\Psi \right) ^{2}+e^{2}\left[ 1+\left(
k_{\phi \phi }\right) _{00}\right] \left[ \left( A_{0}\right) ^{2}+\Psi ^{2}%
\right] \left\vert \phi \right\vert ^{2},  \notag
\end{eqnarray}%
which is positive definite as long as the matrix $\delta _{jk}-\left(
k_{\phi \phi }\right) _{jk}$ is positive definite and $\left( k_{\phi \phi
}\right) _{00}>-1$.

In a similar way as it is done in MCSH model \cite{CSV1,CSV2}, the next step
in the BPS procedure is to impose the condition ${\Psi =\mp {A}_{0},}$ in
the energy density (\ref{ED1}), which becomes
\begin{eqnarray}
\mathcal{E} &=&\frac{1}{2}B^{2}+\frac{1}{2}\left[ ev^{2}-e\eta \left\vert
\phi \right\vert ^{2}\pm \left( k_{AF}\right) _{3}A_{0}\right] ^{2}  \notag
\\[0.2cm]
&&+\left[ \delta _{jk}-\left( k_{\phi \phi }\right) _{jk}\right] \left(
D_{j}\phi \right) ^{\ast }\left( D_{k}\phi \right)  \label{Hstat} \\[0.2cm]
&&+\left( \partial _{k}A_{0}\right) ^{2}+2e^{2}\left[ 1+\left( k_{\phi \phi
}\right) _{00}\right] \left( A_{0}\right) ^{2}\left\vert \phi \right\vert
^{2}.  \notag
\end{eqnarray}

We now introduce a modified vortex ansatz \cite{Carlison1} which considers
Lorentz-violating contributions
\begin{eqnarray}
\phi &=&\frac{v}{\sqrt{\eta }}g\exp \left( i\theta \frac{n}{\Lambda }\right)
,  \notag \\[-0.2cm]
&&  \label{newaz} \\
A_{\theta } &=&-\frac{1}{er}\left( a-\frac{n}{\Lambda }\right) \ \text{\ },\
\text{\ }A_{0}=\omega \left( r\right) ,  \notag
\end{eqnarray}%
where $n$ is called the winding number, which expresses the topological
character of fields. The parameter $\Lambda $ depends only on the
parity-even components of the matrix ruling LV in the Higgs sector:%
\begin{equation}
\Lambda =\sqrt{\frac{1-\left( k_{\phi \phi }\right) _{\theta \theta }}{%
1-\left( k_{\phi \phi }\right) _{rr}}}.
\end{equation}%
The scalar functions $g\left( r\right) $, $a\left( r\right) $ and $\omega
\left( r\right) $ are regular, satisfying the boundary conditions
\begin{eqnarray}
g\left( 0\right) &=&0\ \ ,\ \ a\left( 0\right) =\frac{n}{\Lambda }\ \ ,\ \
\omega ^{\prime }\left( 0\right) =0~,  \notag \\[-0.2cm]
&&  \label{BC} \\
g\left( \infty \right) &=&1\ \ ,\ \ a\left( \infty \right) =0\ \ ,\ \omega
\left( \infty \right) =0~,  \notag
\end{eqnarray}%
these will be explicitly established below in Sec. \ref{BC-cfj}.

Under (\ref{newaz}), the Gauss law is now written as%
\begin{equation}
\omega ^{\prime \prime }+\frac{\omega ^{\prime }}{r}-\left( k_{AF}\right)
_{3}B=2e^{2}v^{2}\Lambda \Delta \omega g^{2},  \label{gauss2}
\end{equation}%
where the magnetic field is $B(r)=\displaystyle-\frac{a^{\prime }}{%
er}$, and the Lorentz-violating parameter $\Delta $ is
given by%
\begin{equation}
\Delta =\frac{1+\left( k_{\phi \phi }\right) _{00}}{\Lambda \eta }>0\mathbf{.%
}
\end{equation}

By replacing the ansatz (\ref{newaz}) in (\ref{Hstat}) and by using Eq.
(\ref{gauss2}) the energy density becomes written as a sum of square terms
plus one term proportional to the magnetic field and a total divergence,
\begin{eqnarray}
\mathcal{E} &=&\frac{1}{2}\left( \frac{{}}{{}}B\mp \left[ ev^{2}\left(
1-g^{2}\right) \pm \left( k_{AF}\right) _{3}\omega \right] \right) ^{2} \\%
[0.2cm]
&&\hspace{-0.5cm}+\frac{v^{2}}{\Lambda }\left( g^{\prime }\mp \Lambda \frac{%
ag}{r}\right) ^{2}\pm ev^{2}B\pm v^{2}\frac{\left( ag^{2}\right) ^{\prime }}{%
r}+\frac{\left( r\omega \omega ^{\prime }\right) ^{\prime }}{r}.  \notag
\end{eqnarray}

The energy will be minimized by equating quadratic terms to zero, obtaining
the self-dual equations
\begin{equation}
g^{\prime }=\pm \Lambda \frac{ag}{r},  \label{AD}
\end{equation}%
\vspace{-0.5cm}
\begin{equation}
B=-\frac{a^{\prime }}{er}=\pm ev^{2}\left( 1-g^{2}\right) +\left(
k_{AF}\right) _{3}\omega ,  \label{B}
\end{equation}%
both together Gauss's law in Eq. (\ref{gauss2}) describe electrically
charged configurations which will be referred as CPT-odd vortices. The first
of these equations, (\ref{AD}), has modifications due to LV present in the
Higgs sector.

Under BPS equations, the energy density reads
\begin{equation}
\mathcal{E}_{BPS}=\pm ev^{2}B\pm v^{2}\frac{\left( ag^{2}\right) ^{\prime }}{%
r}+\frac{\left( r\omega \omega ^{\prime }\right) ^{\prime }}{r}~,
\end{equation}%
whose integration under boundary conditions (\ref{BC}) provides the total
BPS energy%
\begin{equation}
E_{BPS}=\pm ev^{2}\int d^{2}r~B=\pm ev^{2}\Phi =\pm 2\pi v^{2}\frac{n}{%
\Lambda }\ ,  \label{flux_cfj}
\end{equation}%
which remains proportional to the magnetic flux, which, besides being
proportional to the winding number, also depends explicitly on LV belonging
to the Higgs sector.

We use BPS equations and Gauss's law to express energy density as a sum of
positive terms:%
\begin{equation}
\mathcal{E}_{BPS}=B^{2}+2\Lambda v^{2}\left( \frac{ag}{r}\right)
^{2}+2e^{2}v^{2}\Lambda \Delta \left( \omega g\right) ^{2}+\left( \omega
^{\prime }\right) ^{2}~\text{,}
\end{equation}%
it is positive definite because $\Lambda ,\Delta >0.$

Also from Gauss's law in (\ref{gauss2}), the total charge of the self-dual
vortices is%
\begin{equation}
Q=2\pi \int_{0}^{\infty }dr~r~q_{0}=\frac{\left( k_{AF}\right) _{3}}{e}\Phi
=2\pi \frac{\left( k_{AF}\right) _{3}}{e}\frac{n}{\Lambda },
\label{charge_cfj}
\end{equation}%
where $q_{0}=-2ev^{2}\Lambda \Delta \omega g^{2},$ is the electric charge
density. The result is that the total electric charge is proportional to $%
n/\Lambda $\ just as the magnetic flux is.

\subsection{Asymptotic behavior\label{BC-cfj}}

The asymptotic behavior is studied by solving BPS equations (\ref{AD}) and (\ref%
{B}) together with Gauss's law (\ref{gauss2}) at the limits $r\rightarrow0$
and $r\rightarrow\infty$.

At the origin, the field profiles has the following behavior:
\begin{eqnarray}
g\left( r\right) &=&Gr^{n}+~...~,  \label{g(0)} \\[0.2cm]
a\left( r\right) &=&\frac{n}{\Lambda }-\frac{1}{2}[ ev^{2} +(k_{AF})
_{3}\omega _{0}] er^{2}+~...~,  \label{a(0)} \\[0.2cm]
\omega \left( r\right) &=&\omega _{0}+\frac{1}{4}[ ev^{2}+( k_{AF})\omega
_{0}] ( k_{AF})_{3}r^{2}+...,  \label{A(0)}
\end{eqnarray}%
where $\omega _{0}=\omega \left( 0\right) $. Equation (\ref%
{a(0)}) describes the behavior of $a\left( r\right) $ close to origin, and it
justifies the new ansatz imposed in Eq. (\ref{newaz}). On the other hand,
Eq. (\ref{A(0)}) shows explicitly that the electric field at the origin must
be null, $\omega ^{\prime }\left( 0\right) =0$.

It can be shown this model supports ANO vortex solutions whose
characteristic behavior, at $r\rightarrow \infty $, for field profiles is
\begin{eqnarray}
1-g\left( r\right) &\sim &r^{-1/2}e^{-\beta r}\ \text{,}\   \notag \\[0.08in]
a\left( r\right) &\sim &r^{1/2}e^{-\beta r}\ \text{,}\   \label{anocc} \\%
[0.08in]
\omega \left( r\right) &\sim &r^{-1/2}e^{-\beta r}~\text{,}  \notag
\end{eqnarray}%
where $\beta $ is a real and positive parameter related to the spatial
extension of the vortex, given by%
\begin{eqnarray}
\beta &=&\frac{1}{2}\sqrt{\left[ \left( k_{AF}\right) _{3}\right]
^{2}+2e^{2}v^{2}\Lambda \left( \sqrt{\Delta }+1\right) ^{2}}  \label{BETA} \\%
[0.2cm]
&&-\frac{1}{2}\sqrt{\left[ \left( k_{AF}\right) _{3}\right]
^{2}+2e^{2}v^{2}\Lambda \left( \sqrt{\Delta }-1\right) ^{2}}.  \notag
\end{eqnarray}%
This shows that for fixed $\left( k_{AF}\right) _{3}$, the asymptotic
behavior of the vortex profiles is ruled by the Lorentz-violating parameters belonging to
the Higgs sector. Under such circumstance we have two interesting
situations: The first one occurs when $\Delta =1$, $\Lambda =1$, and $\beta $
behaves as
\begin{equation}
\beta \rightarrow \frac{1}{2}\sqrt{\left[ \left( k_{AF}\right) _{3}\right]
^{2}+8e^{2}v^{2}}-\frac{1}{2}\left\vert \left( k_{AF}\right) _{3}\right\vert
.
\end{equation}%
Such a value is exactly the same ruling the asymptotic behavior of MCSH
vortices. The second interesting case occurs when $\Delta $ takes
sufficiently large values ($\Delta \gg \left( k_{AF}\right) _{3}$):
\begin{equation}
\beta \rightarrow ev\Lambda ^{1/2}\sqrt{2},
\end{equation}%
This value corresponds to the usual MH model with Lorentz violation only in
the Higgs sector, and it can be verified explicitly in Ref. \cite{Carlison1}.
On the other hand, by fixing the Lorentz-violating parameters of the Higgs sector and
sufficiently large values of $\left( k_{AF}\right) _{3}$, $\beta $ behaves as%
\begin{equation}
\beta \rightarrow \frac{2e^{2}v^{2}}{\left( k_{AF}\right) _{3}}\Lambda \sqrt{%
\Delta }~,
\end{equation}%
which is the mass scale of a CSH model with Lorentz-violating terms only in the Higgs
sector.

\subsection{Numerical analysis}

In order to perform numerical analysis, we use dimensionless coordinates and
fields. We define the dimensionless coordinate $\rho =ev\Lambda ^{1/2}r$ and
perform the following field redefinitions:%
\begin{eqnarray}
& \displaystyle{g\left( r\right) \rightarrow g\left( \rho \right)
\;,\;\;a\left( r\right) \rightarrow \frac{a\left( \rho \right) }{\Lambda }%
\;,\;\;\omega \left( r\right) \rightarrow \frac{v}{\Lambda ^{1/2}}\omega
\left( \rho \right) ,} &  \notag \\[0.06in]
&\displaystyle{B\left( r\right) \rightarrow ev^{2}B\left( \rho \right)
\;,\;\;\mathcal{E}_{BPS}\left( r\right) \rightarrow \frac{v^{2}}{\Lambda }%
\mathcal{E}_{BPS}\left( \rho \right) ,}&  \label{redef} \\[0.08in]
&\displaystyle{\left( k_{AF}\right) _{3}\rightarrow ev\Lambda ^{1/2}\kappa .}%
&  \notag
\end{eqnarray}

With these new definitions, we rewrite the BPS equations and Gauss's law:
\begin{eqnarray}
&\displaystyle{g^{\prime }=\pm \frac{ag}{\rho }~,}&  \label{xxq1} \\[0.2cm]
&\displaystyle{-\frac{a^{\prime }}{\rho }=\pm \left( 1-g^{2}\right) +\kappa
\omega ~,}&  \label{xxq2} \\
&\displaystyle{\omega ^{\prime \prime }+\frac{\omega ^{\prime }}{\rho }%
+\kappa \frac{a^{\prime }}{\rho }=2\Delta \omega g^{2}~\text{.}}&
\label{xxq3}
\end{eqnarray}%
It can be observed that if $\Delta =1$, Eqs. (\ref{xxq1}-\ref{xxq3}) looks
exactly like the ones in MCSH model. The set of equations will be solved by
using the following boundary condition:%
\begin{eqnarray}
g\left( 0\right) &=&0\ \text{,}\ a\left( 0\right) =n\ \text{,}\ \omega
^{\prime }\left( 0\right) =0~\text{,} \\[0.2cm]
g\left( \infty \right) &=&1\ \text{,}\ a\left( \infty \right) =0\ \text{,}\
\omega \left( \infty \right) =0.  \label{bccfj}
\end{eqnarray}

Also the BPS energy density is expressed by
\begin{equation}
\mathcal{E}_{BPS}\left( \rho \right) =B^{2}+2\left( \frac{ag}{\rho }\right)
^{2}+2\Delta \left( \omega g\right) ^{2}+\left( \omega ^{\prime }\right)
^{2};  \label{defen}
\end{equation}
it will be positive definite whenever $\Delta>0$.

The dimensionless $\beta $ mass scale reads
\begin{eqnarray}
\beta &=&\frac{1}{2}\sqrt{\kappa ^{2}+2\left( \sqrt{\Delta }+1\right) ^{2}}
\label{BETA1} \\[0.15cm]
&&-\frac{1}{2}\sqrt{\kappa ^{2}+2\left( \sqrt{\Delta }-1\right) ^{2}}.
\notag
\end{eqnarray}

Below, we depict the profiles obtained from numerical solutions of Eqs. (\ref%
{xxq1}--\ref{xxq3}) under boundary conditions (\ref{bccfj}) for $\kappa =1$,
$n=1$ and some values for $\Delta $. Because the BPS energy density (\ref%
{defen}) is positive definite for $\Delta >0$, we consider two regions $%
0<\Delta <1$ (green lines) and $\Delta >1$ (red lines), in which the behavior
of solutions are different. There are two interesting values of $\Delta $
which allow to obtain two well-know models: The first one is $\Delta =1$
(solid black lines) whose profiles correspond to MCSH model, and the second
value is the limit $\Delta \rightarrow \infty ,$ when MH model (solid blue
lines) is attained.

\begin{figure}[H]
\centering\includegraphics[width=8.25cm]{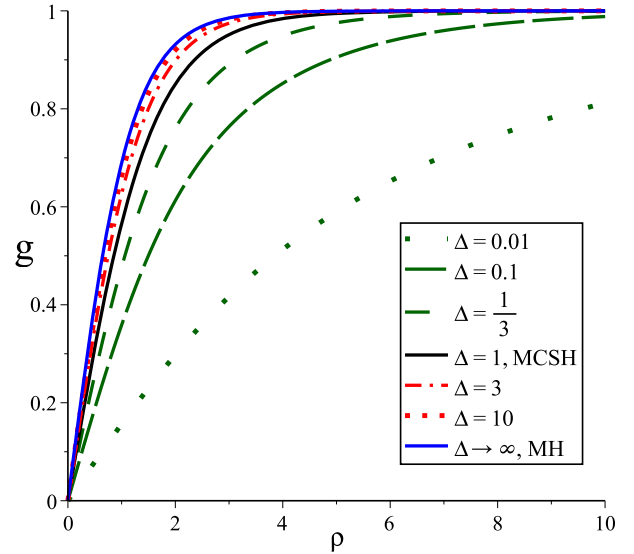}
\caption{Scalar field $g(\protect\rho )$.}
\label{S_BPSx}
\end{figure}

\begin{figure}[H]
\centering
\includegraphics[width=8.25cm]{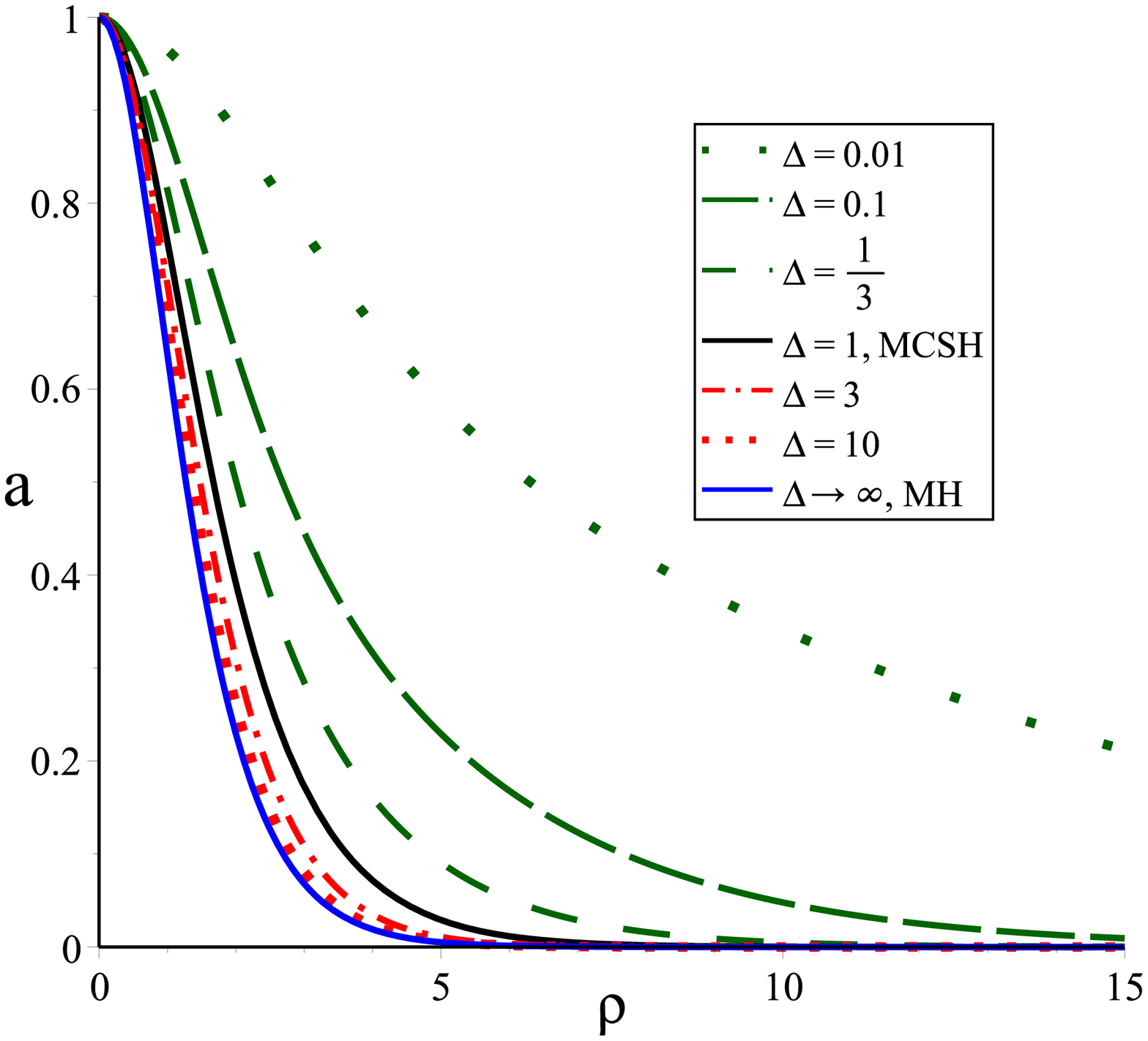}
\caption{Vector potential ${a}(\protect\rho )$. }
\label{A_BPS}
\end{figure}

Figures \ref{S_BPSx} and \ref{A_BPS} depict the profiles of the Higgs and vector
fields, respectively. It is observed that they reach the asymptotic values
rapidly for $\Delta >1$ and saturate more slowly for $\Delta <1,$ as is
expected from Eq. (\ref{BETA1}) for fixed $\kappa $. Also, it is verified
explicitly that the solutions for $1<\Delta <\infty $ (red lines) are
confined between the MCSH (solid black line) and MH (solid blue line)
models. Such that the profiles are narrower when $\Delta >1$, and the
maximum narrowing is attained when $\Delta \rightarrow \infty $ (solid blue
line) where the MH model is recovered. On the other hand, for $0<\Delta <1$, the
profiles get more spread out, occupying a larger area when $\Delta
\rightarrow 0$. In this limit, the vector potential has a very light decay,
by which we can infer that the magnetic field corresponding to this values
will have low intensity.

\begin{figure}[H]
\centering\includegraphics[width=8.25cm]{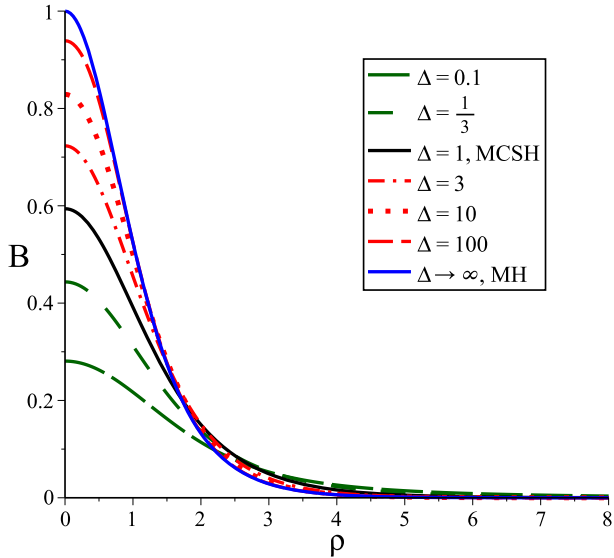}
\caption{Magnetic field ${B}(\protect\rho )$.}
\label{B_BPS}
\end{figure}

Figure \ref{B_BPS} shows the profiles for the magnetic field which are lumps
centered at the origin. For increasing values of $\Delta $, the profiles are
narrower and have higher amplitude in such a way the maximum narrowing and
maximum intensity are reached when $\Delta \rightarrow \infty $ (solid blue
line). For $0<\Delta <1$, the profiles have less intensity and are more
spread when $\Delta \rightarrow 0$. Higher amplitudes are obtained for the
values $1<\Delta <\infty $; such profiles are located between the MCSH ($%
\Delta =1$, solid black line) and the MH profile ($\Delta \rightarrow \infty
$, solid blue line). At infinity, the magnetic field vanishes more rapidly as
$\Delta $ increases.

\begin{figure}[H]
\centering
\includegraphics[width=8.25cm]{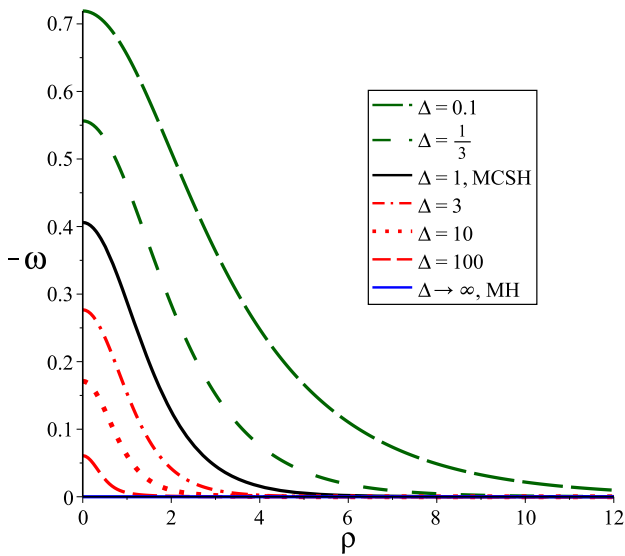}
\caption{Scalar potential $\protect\omega (\protect\rho )$.}
\label{W_BPS}
\end{figure}

\begin{figure}[H]
\centering
\includegraphics[width=8.25cm]{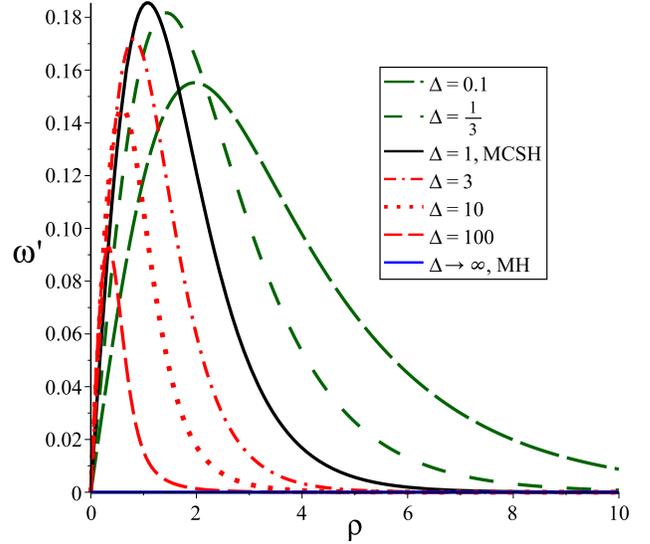}
\caption{Electric field $\protect\omega ^{\prime }(\protect\rho )$.}
\label{El_BPS}
\end{figure}

Figure \ref{W_BPS} depicts the profiles for the scalar potential. For $\Delta
>1$, the CFJ parameter is less relevant for increasing values of $\Delta $,
such that the profiles become smaller, and in the limit $\Delta \rightarrow
\infty $ (solid blue line), the profile overlaps the horizontal axis, implying
that vortices become electrically uncharged such as in the MH model. On the
other hand, in the interval $0<\Delta <1$, the scalar potential profiles are
more extended and with greater intensity at the origin. For this range of $%
\Delta $, the CFJ parameter becomes more relevant when $\Delta \rightarrow 0$.

\begin{figure}[H]
\centering
\includegraphics[width=8.25cm]{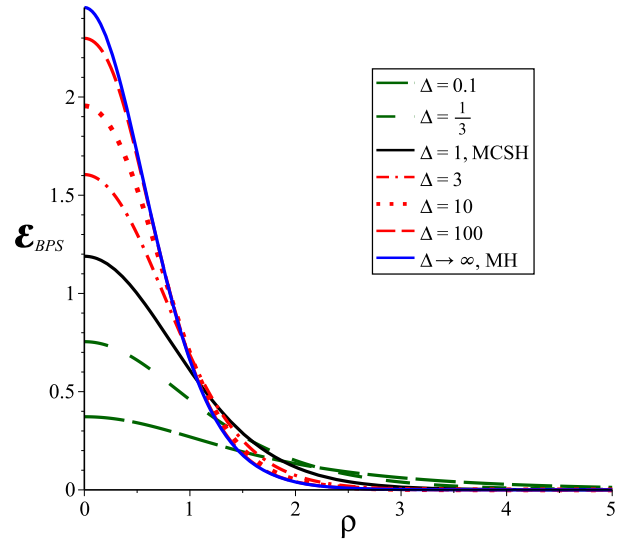}
\caption{Energy density $\mathcal{E}_{BPS}(\protect\rho )$.}
\label{Energy_BPS}
\end{figure}

Figure \ref{El_BPS} describes the behavior for the electric field. For $n=1$,
the maximum electric field intensity is reached for $\Delta =1$ (solid black
line), corresponding to MCSH model. The electric field gets radially spread out
for decreasing values of $\Delta $. On the other hand, for $\Delta >1,$\ the
electric field is located closer to the origin and decays rapidly for
increasing values of $\Delta $. In the limit $\Delta \rightarrow \infty ,$
the electric field disappears, which result agrees with electrically
uncharged vortex models.

Figure \ref{Energy_BPS} describes BPS energy density profiles, which for $n=1$
are lumps centered at origin. For increasing values of $\Delta ,$\ we have
higher amplitudes at the origin which sharply decay, reaching zero rapidly, which
corresponds to well-located vortices. On the other hand, when $\Delta
\rightarrow 0$, amplitude decreases at the origin and reaches zero slowly,
implying a more extended, large energy density distribution but with
little intensity. The profiles of BPS energy density have the same behavior as the magnetic field profiles.

\section{TOPOLOGICAL CHARGED BPS VORTICES IN A CPT-EVEN AND LORENTZ-VIOLATING FRAMEWORK}

In this section, we investigate the formation of topological
electrically charged BPS vortices in a Lorentz-violating MH model, where
both the gauge and Higgs sector contain CPT-even and Lorentz-violating terms
\cite{Colladay2}. Specifically, we analyze the nonbirefringent Lorentz violating
coefficients of the electromagnetic sector of SME. This model is a natural
generalization of earlier studies performed in Refs.\cite{Carlison1,Carlison2}.
The Lagrangian density for this model describing charged vortices is
\begin{eqnarray}
\mathcal{L} &=&-\frac{1}{4}F_{\mu \nu }F^{\mu \nu }-\frac{1}{2}\kappa ^{\rho
\alpha }F_{\rho \sigma }F_{\alpha }^{\text{ \ }\sigma }+\left\vert D_{\mu
}\phi \right\vert ^{2}  \notag \\[0.2cm]
&&+\left( k_{\phi \phi }\right) ^{\mu \nu }\left( D_{\mu }\phi \right)
^{\ast }\left( D_{\nu }\phi \right) -U\left( \left\vert \phi \right\vert
\right) ~\text{.}  \label{Lag2}
\end{eqnarray}%
In stationary regime, Gauss's and Ampere's law are %
\begin{eqnarray}
L_{ij}\partial _{i}\partial _{j}A_{0}+\epsilon _{ija}\kappa _{0i}\partial
_{j}B_{a} &=&-e\mathcal{J}_{0}~\text{,} \\[0.2cm]
{M}_{jbc}\partial _{b}B_{c}-\kappa _{0i}\partial _{j}\partial
_{j}A_{0}+\kappa _{0j}\partial _{j}\partial _{i}A_{0} &=&e\mathcal{J}_{i}~%
\text{,}
\end{eqnarray}%
respectively. The matrix $L_{ij}=\left( 1+\kappa _{00}\right)
\delta _{ij}-\kappa _{ij}$ is symmetric and positive definite
which carries parity-even and CPT-even Lorentz-violating coefficients. The density $%
\mathcal{J}_{0}$ is the same defined after Eq. (\ref{gauss}). In
Ampere's law, we have defined the tensor $M_{jbc}=\epsilon _{jbc}-\epsilon
_{jac\ }\kappa _{ab}-\kappa _{ja}\epsilon _{abc}$, and current
density $\mathcal{J}_{i}=J_{i}-\left( k_{\phi \phi }\right)
_{ij}J_{j}+\left( k_{\phi \phi }\right) _{0i}J_{0}$, with $J_{\mu }$%
 also defined after Eq. (\ref{gauss}). We observe that parity-odd
coefficients $\kappa _{0i}$ and $\left( k_{\phi \phi
}\right) _{0i}$ are responsible for electric and magnetic sector
coupling however some $\left( k_{\phi \phi }\right) _{0i}\ $ do not
contribute to the formation of vortex solutions.

As was shown in Ref. \cite{Carlison2}, the model (\ref{Lag2}) must be
modified to supports BPS solutions:
\begin{eqnarray}
\mathcal{L} &=&\mathcal{-}\frac{1}{4}F_{\mu \nu }F^{\mu \nu }-\frac{1}{2}%
\kappa ^{\rho \alpha }F_{\rho \sigma }F_{\alpha }{}^{\sigma }  \notag \\%
[0.2cm]
&&+\left\vert D_{\mu }\phi \right\vert ^{2}+\left( k_{\phi \phi }\right)
^{\mu \nu }\left( D_{\mu }\phi \right) ^{\ast }\left( D_{\nu }\phi \right)
\notag \\[-0.2cm]
&&  \label{Lag MHpara} \\
&&+\frac{1}{2}\left( 1+\kappa _{00}\right) \partial _{\mu }\Psi \partial
^{\mu }\Psi +\frac{1}{2}\kappa ^{\mu \nu }\partial _{\mu }\Psi \partial
_{\nu }\Psi  \notag \\[0.2cm]
&&-e^{2}\left[ 1+\left( k_{\phi \phi }\right) _{00}\right] \Psi
^{2}\left\vert \phi \right\vert ^{2}-U\left( \left\vert \phi \right\vert
,\Psi \right) ~\text{.}  \notag
\end{eqnarray}

The interaction term $U\left( \left\vert \phi \right\vert ,\Psi \right) $
given by
\begin{equation}
U\left( \left\vert \phi \right\vert ,\Psi \right) =\frac{\left( ev^{2}-e\eta
\left\vert \phi \right\vert ^{2}-\epsilon _{ij}\kappa _{0i}\partial _{j}\Psi
\right) ^{2}}{2\left( 1-s\right) }  \label{P}
\end{equation}%
contains a derivative coupling and a $\left\vert \phi \right\vert ^{4}-$%
potential, which provides BPS configurations. Above we have define the
parameter $s=\kappa _{11}+\kappa _{22}=\kappa _{rr}+\kappa _{\theta \theta }$%
.

\subsection{BPS formalism}

In order to implement the BPS formalism, we first write down the stationary
energy density%
\begin{eqnarray}
\mathcal{E} &=&\frac{1}{2}L_{ij}\left( \partial _{i}A_{0}\right) \left(
\partial _{j}A_{0}\right) +\frac{1}{2}\left( 1-s\right) B^{2}  \notag \\%
[0.08in]
&&\hspace{-0.7cm}+\left[ \delta _{ij}-\left( k_{\phi \phi }\right) _{ij}%
\right] \left( D_{i}\phi \right) ^{\ast }\left( D_{j}\phi \right) +\frac{1}{2%
}L_{ij}\partial _{i}\Psi \partial _{j}\Psi  \label{E_Xa} \\[0.08in]
&&\hspace{-0.7cm}+e^{2}\left[ 1+\left( k_{\phi \phi }\right) _{00}\right] %
\left[ \left( A_{0}\right) ^{2}+\Psi ^{2}\right] \left\vert \phi \right\vert
^{2}+U\left( \left\vert \phi \right\vert ,\Psi \right) .  \notag
\end{eqnarray}%
It is positive-definite whenever $s<1$,$\ \left( k_{\phi \phi }\right)
_{00}>-1$, and $\delta _{ij}-\left( k_{\phi \phi }\right) _{ij},$ a
positive-definite matrix.

By following a similar procedure to that used in the previous model, we
impose condition $\Psi =\mp A_{0},$ in the energy density (\ref%
{E_Xa}), which becomes%
\begin{eqnarray}
\mathcal{E} &=&\frac{1}{2}\left( 1-s\right) B^{2}+\frac{\left( ev^{2}-e\eta
\left\vert \phi \right\vert ^{2}\pm \epsilon _{ij}\kappa _{0i}\partial
_{j}A_{0}\right) ^{2}}{2\left( 1-s\right) }  \notag \\[0.08in]
&&\hspace{-0.7cm}+\left[ \delta _{ij}-\left( k_{\phi \phi }\right) _{ij}%
\right] \left( D_{i}\phi \right) ^{\ast }\left( D_{j}\phi \right)
\label{E_X} \\[0.08in]
&&\hspace{-0.7cm}+L_{ij}\left( \partial _{i}A_{0}\right) \left( \partial
_{j}A_{0}\right) +2e^{2}\left[ 1+\left( k_{\phi \phi }\right) _{00}\right]
\left( A_{0}\right) ^{2}\left\vert \phi \right\vert ^{2}.  \notag
\end{eqnarray}

Now we introduce the vortex ansatz (\ref{newaz}) to express Gauss's
law as follows:%
\begin{equation}
{\left( 1+\lambda _{r}\right) }\left( \omega ^{\prime \prime }+\frac{\omega
^{\prime }}{r}\right) -\kappa _{0\theta }\frac{\left( r{B}\right) ^{\prime }%
}{r}={2e}^{2}v^{2}\Lambda \Delta {\omega g^{2},}  \label{GD_3}
\end{equation}%
with $\lambda _{r}=\kappa _{00}-\kappa _{rr}$. Afterwards
rewriting the energy density (\ref{E_X}) by using the ansatz (\ref{newaz}), we
use Eq. (\ref{GD_3}) to express it in the following convenient form:%
\begin{eqnarray}
\mathcal{E} &=&\frac{1}{2}\left( 1-s\right) \left[ B\mp \frac{ev^{2}\left(
1-g^{2}\right) \mp \kappa _{0\theta }\omega ^{\prime }}{1-s}\right] ^{2}
\notag \\[0.2cm]
&&+\frac{v^{2}}{\Lambda }\left[ ~g^{\prime }\mp \Lambda \frac{ag}{r}\right]
^{2}\pm ev^{2}B\pm v^{2}\frac{\left( ag^{2}\right) ^{\prime }}{r}  \label{E2}
\\[0.2cm]
&&+(1+\lambda _{r})\frac{\left( r\omega \omega ^{\prime }\right) ^{\prime }}{%
r}-\kappa _{0\theta }\frac{\left( r\omega B\right) ^{\prime }}{r}.  \notag
\end{eqnarray}%
It is minimized by equating quadratic terms to zero, providing self-dual or
BPS equations
\begin{eqnarray}
&\displaystyle{g^{\prime }=\pm \Lambda \frac{ag}{r}~,}&  \label{GD_1} \\%
[0.2cm]
&\displaystyle{B=-\frac{a^{\prime }}{er}=\pm \frac{ev^{2}\left(
1-g^{2}\right) }{1-s}-\frac{\kappa _{0\theta }\omega ^{\prime }}{1-s}~}&,
\label{GD_2}
\end{eqnarray}%
both together with Gauss's law in Eq. (\ref{GD_3}) describe electrically neutral
configurations, which will be called as CPT-even vortices. The total
electric charge of the topological vortices is zero, as can explicitly
shown by integrating Gauss's law in Eq. (\ref{GD_3}) under boundary
conditions given in Eqs. (\ref{BCGD1}) and (\ref{BCGD2}):
\begin{equation}
Q={4\pi e}v^{2}\Lambda \Delta \int_{0}^{\infty }dr~r{\omega g^{2}=0.}
\end{equation}

Whenever the BPS equations are satisfy, the BPS energy density is
\begin{eqnarray}
\mathcal{E}_{BPS} &=&\pm ev^{2}B\pm v^{2}\frac{\left( ag^{2}\right) ^{\prime
}}{r}  \label{Eden} \\[0.2cm]
&&+(1+\lambda _{r})\frac{\left( r\omega \omega ^{\prime }\right) ^{\prime }}{%
r}-\kappa _{0\theta }\frac{\left( r\omega B\right) ^{\prime }}{r},  \notag
\end{eqnarray}%
whose integration under boundary conditions established in Eqs. (\ref{BCGD1}%
) and (\ref{BCGD2}) provides total BPS\ energy
\begin{equation}
E_{BPS}=\pm ev^{2}\int Bd^{2}x=\pm 2\pi v^{2}\frac{n}{\Lambda }~,
\label{flux_mhLV}
\end{equation}%
where the magnetic flux, such as in the previous model, besides being
proportional to the winding number, also depends explicitly on LV belonging
to the Higgs sector.

We use BPS equations and Gauss's law to express energy density as
\begin{eqnarray}
\mathcal{E}_{BPS} &=&\left( 1-s\right) B^{2}+2\Lambda v^{2}\left( \frac{ag}{r%
}\right) ^{2}  \notag \\[-0.2cm]
&& \\
&&+2e^{2}v^{2}\Lambda \Delta \left( g\omega \right) ^{2}+\left( 1+\lambda
_{r}\right) \left( \omega ^{\prime }\right) ^{2};  \notag
\end{eqnarray}%
it is positive-definite for $s<1$ and $\lambda _{r}>-1$.

\subsection{Boundary conditions}

By solving BPS\ equations\ (\ref{GD_1} and \ref{GD_2}) and Gauss's law (\ref%
{GD_3}) at $r=0$, we obtain
\begin{eqnarray}
g\left( r\right) &=&G_{n}r^{n}+\mathcal{\ldots },  \label{gg(r)} \\[0.2cm]
a\left( r\right) &=&{\frac{n}{\Lambda }}-\,{\frac{{e}^{2}{v}^{2}\left(
1+\lambda _{r}\right) ~}{2\Sigma }r}^{2}+\mathcal{\ldots },  \label{aa(r)} \\%
[0.2cm]
\omega \left( r\right) &=&\omega _{0}+{\frac{e{v}^{2}\kappa _{0\theta }}{%
\Sigma }}~r+\mathcal{\ldots },  \label{ww(r)}
\end{eqnarray}%
where $\omega _{0}=\omega \left( 0\right) $ and %
\begin{equation}
\Sigma =\left( \kappa _{0\theta }\right) ^{2}+\left( 1-s\right) \left(
1+\lambda _{r}\right).  \label{sigma}
\end{equation}%
Therefore, the boundary conditions at the origin are %
\begin{equation}
g\left( 0\right) =0,~a\left( 0\right) =\frac{n}{\Lambda },~\omega ^{\prime
}\left( 0\right) =\frac{e{v}^{2}\kappa _{0\theta }}{\Sigma }  \label{BCGD1}
\end{equation}

The model also supports ANO vortices, whose behavior when $r\rightarrow
\infty $ is given by Eq. (\ref{anocc}), so the boundary conditions for the
field profiles at infinity are
\begin{equation}
g\left( \infty \right) =1\ ,\ \ a\left( \infty \right) =0\ ,\ \ \omega
\left( \infty \right) =0,  \label{BCGD2}
\end{equation}%
but now the mass scale $\beta $ is
\begin{equation}
\beta =ev\Lambda ^{1/2}\sqrt{\frac{\beta _{1}\pm \beta _{2}}{\Sigma }},
\label{betax}
\end{equation}%
where
\begin{eqnarray}
\beta _{1} &=&\left( 1+\lambda _{r}\right) +\Delta \left( 1-s\right) , \\%
[0.3cm]
\beta _{2} &=&\sqrt{\left[ \left( 1+\lambda _{r}\right) -\Delta \left(
1-s\right) \right] ^{2}-4\Delta \left( \kappa _{0\theta }\right) ^{2}},
\end{eqnarray}%
and $\beta _{2},\;\beta _{2}$ are positive real numbers. For $\beta _{2}$, the
condition $4\Delta \left( \kappa _{0\theta }\right) ^{2}\leq \left[ \left(
1+\lambda _{r}\right) -\Delta \left( 1-s\right) \right] ^{2}$ must be
satisfied. The signals will be used as follows: $+\left( -\right) $ for $%
\left( 1+\lambda _{r}\right) -\Delta \left( 1-s\right) >0\left( <0\right) $.

When $\kappa _{0\theta }=0$, the Eq. (\ref{betax}) recovers the mass scale
of electrically uncharged BPS vortices,
\begin{equation}
\beta =ev\Lambda ^{1/2}\sqrt{\frac{2}{1-s}},
\end{equation}%
already obtained in Ref. \cite{Carlison1}.

On the other hand, when $\beta _{2}=0$, the parity-odd coefficient can be
expressed in terms of parity-even ones,
\begin{equation}
\kappa _{0\theta }=\pm\frac{\left\vert 1+\lambda _{r}-\Delta \left(
1-s\right) \right\vert }{2\sqrt{\Delta }},
\end{equation}%
and the mass scale becomes%
\begin{equation}
\beta =\ ev\Lambda ^{1/2}\sqrt{\frac{4\Delta }{\left( 1+\lambda _{r}\right)
+\Delta \left( 1-s\right) }}.
\end{equation}

Another interesting possibility is to set $s=0$ and $\lambda _{r}=0$ in Eq.(%
\ref{betax}); i.e., we can consider null parity-even electromagnetic
coefficients, getting
\begin{equation}
\beta =ev\Lambda ^{1/2}\sqrt{\frac{1+\Delta \pm \sqrt{\left( 1-\Delta
\right) ^{2}-4\Delta \left( \kappa _{0\theta }\right) ^{2}}}{1+\left( \kappa
_{0\theta }\right) ^{2}}}\ ,  \label{betax2}
\end{equation}%
with signal $+\left( -\right) $ for $1-\Delta >0\left( <0\right) $,
respectively. This situation also provides ANO-like vortices whenever
\begin{equation}
\left( 1-\Delta \right) ^{2}\geq 4\Delta \left( \kappa _{0\theta }\right)
^{2}
\end{equation}%
is satisfied. We remark that in the absence of LV in Higgs sector ($\Delta
=1 $) (see Ref. \cite{Carlison2}), it is impossible to obtain ANO vortices
when $s$\ and $\lambda _{r}$ are nulls, because the mass scale (\ref{betax2})
becomes a complex number.

\subsection{Numerical analysis}

Under the same coordinate rescaling and field redefinitions as in (\ref%
{redef}) but with $\kappa _{0\theta }\rightarrow \kappa $, the dimensionless
BPS equations and Gauss's law are
\begin{eqnarray}
&\displaystyle{g^{\prime }=\pm \frac{ag}{\rho }~\ ,}&  \label{eqt2a} \\%
[0.2cm]
&\displaystyle{-\frac{a}{\rho }=\pm \frac{\left( 1-g^{2}\right) }{\left(
1-s\right) }-\frac{\kappa }{\left( 1-s\right) }\omega ^{\prime },}&
\label{eqt2b} \\[0.2cm]
&\displaystyle{\left( 1+\lambda _{r}\right) \left( \omega ^{\prime \prime }+%
\frac{\omega ^{\prime }}{\rho }\right) -{\kappa }\frac{\left( \rho {B}%
\right) ^{\prime }}{\rho }=2\Delta g^{2}\omega .}&  \label{eqt2c}
\end{eqnarray}%
We can see that for $\kappa \rightarrow -\kappa $, the solutions change as $%
g\rightarrow g$, $a\rightarrow a$, $\omega \rightarrow -\omega $.

On the other hand, the BPS energy density is given by
\begin{eqnarray}
\mathcal{E}_{_{BPS}}\left( \rho \right) &=&\left( 1-s\right) B^{2}+2\left(
\frac{ag}{\rho }\right) ^{2}  \notag \\[-0.3cm]
&&  \label{H_DEF_POSx} \\
&&+2\Delta \left( {g\omega }\right) ^{2}+\left( 1+\lambda _{r}\right) \left(
\omega ^{\prime }\right) ^{2},  \notag
\end{eqnarray}%
is defined positive providing that $s<1$, $\Delta >0$, $\lambda _{r}>-1$.

\subsubsection{A charged vortex configuration}

Now we consider the following configuration to solve numerically the
dimensionless Eqs. (\ref{eqt2a})-(\ref{eqt2c}):
\begin{equation}
\lambda _{r}=0~\ ,~\ s=0~\ ,~\ \kappa =\frac{\Delta -1}{2\sqrt{\Delta }}
~,~\ \Delta >0 .
\end{equation}
In this case, the boundary conditions are expressed as
\begin{eqnarray}
g\left( 0\right) &=&0,~a\left( 0\right) =n,~{\omega }^{\prime }\left(
0\right) ={\frac{2\left( \Delta -1\right) \sqrt{\Delta }}{\left( 1+\Delta
\right) ^{2}}},  \notag \\[-0.15cm]
&&  \label{bcsxx} \\[-0.15cm]
g\left( \infty \right) &=&1,~a\left( \infty \right) =0,~{\omega }\left(
\infty \right) =0.  \notag
\end{eqnarray}%
The mass scale $\beta $ is given by%
\begin{equation}
\beta =2\sqrt{\frac{\Delta }{\left( 1+\Delta \right) }},  \label{betaMH}
\end{equation}%
and we note that for $\Delta \ll 1$, the defect reaches its asymptotic
values slowly. But when $\Delta \sim 1$, the behavior is close to the MH
profiles (see below, Figs. \ref{HIGGS}--\ref{ED}, solid black lines), because
$\beta \rightarrow \sqrt{2}$, MH mass scale. On the other hand, for $\Delta
\rightarrow \infty $, the mass scale reaches its maximum value $\beta
\rightarrow 2$ (see below, Figs. \ref{HIGGS}-\ref{MF},\ref{EF},\ref{ED},
solid blue lines).

Below, we depict the profiles obtained from numerical solutions of the Eqs. (%
\ref{eqt2a})-(\ref{eqt2c}) under boundary conditions (\ref{bcsxx}). Without
loss of generality, we set $n=1$. Because the BPS energy density is
positive definite for $\Delta >0$, we consider two regions $0<\Delta <1$
(green lines) and $\Delta >1$ (red lines) in which the behavior of solutions
are different. The solutions are always compared with profiles of MH model
(solid black lines).

\begin{figure}[H]
\centering\includegraphics[width=4.35cm]{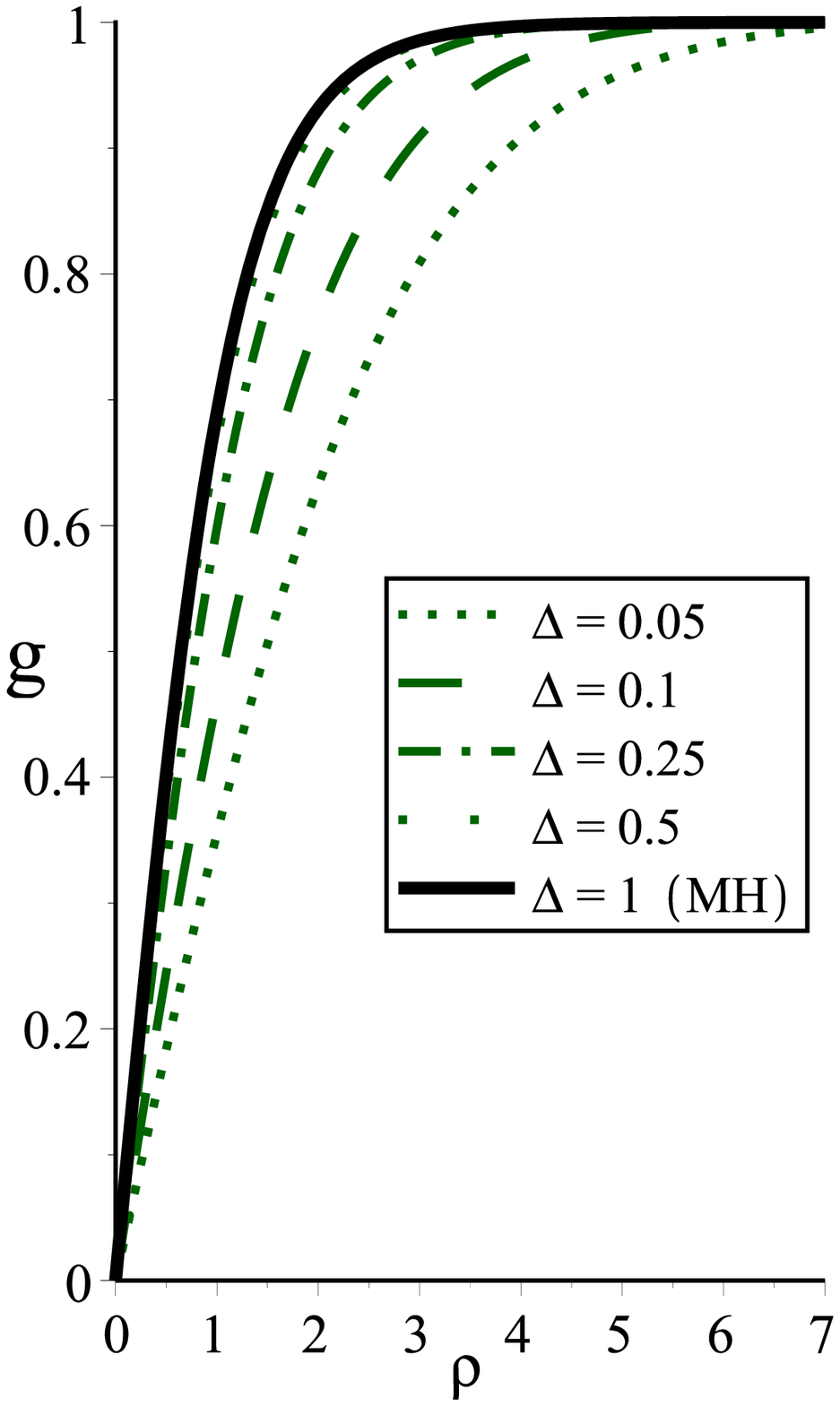}\!\! %
\includegraphics[width=4.25cm]{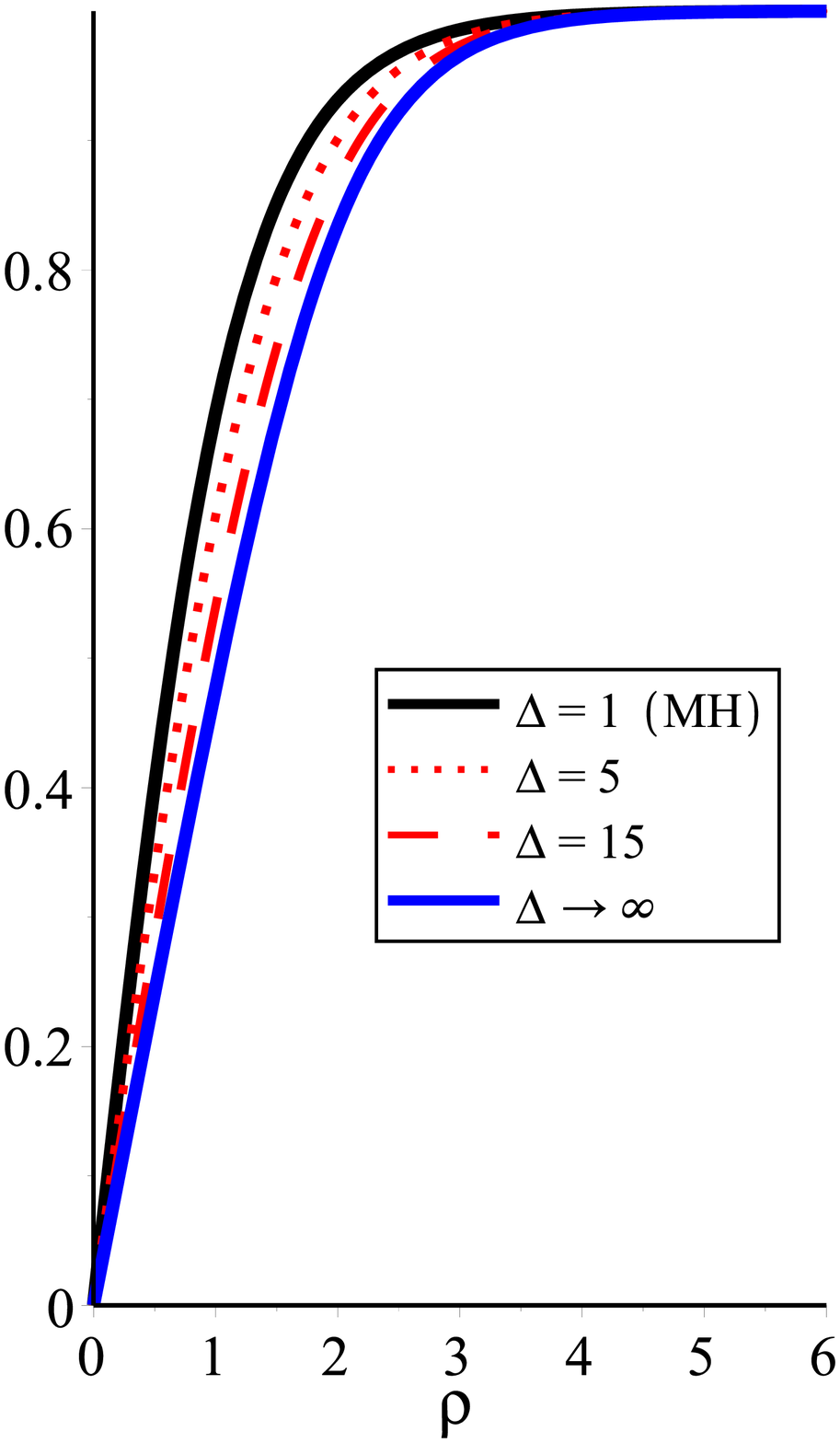}
\caption{ Higgs field ${g}(\protect\rho)$. Left-side represents the profiles
for $0\leq\Delta\leq 1$ and right-side for $\Delta\geq 1$.}
\label{HIGGS}
\end{figure}

Figure \ref{HIGGS} shows the profiles of the Higgs field. For $\Delta \ll 1,$ they
are very spread and reach their asymptotic value slowly. But when $\Delta
\rightarrow 1,$ they are narrower and attain the vacuum state more rapidly.
For $0\leq \Delta \leq 1,$ the profiles are limited by the usual MH solution
$\Delta =1$ (solid black line). For $\Delta >1$, the profiles are wider than
the MH one, but such increment is limited by the profile obtained when $%
\Delta \rightarrow \infty $ (solid blue line).

\begin{figure}[H]
\centering
\includegraphics[width=4.35cm]{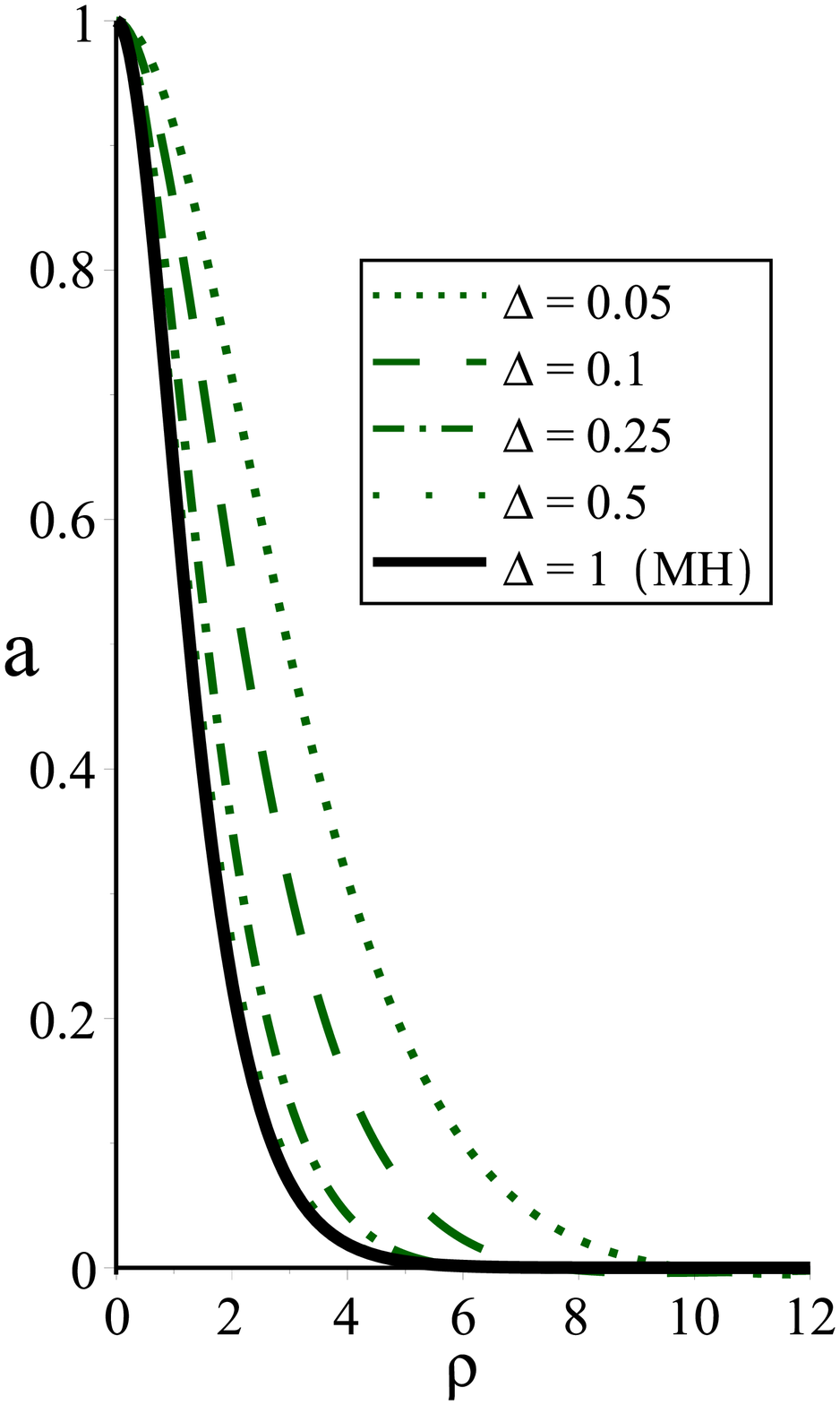}\!\!%
\includegraphics[width=4.25cm]{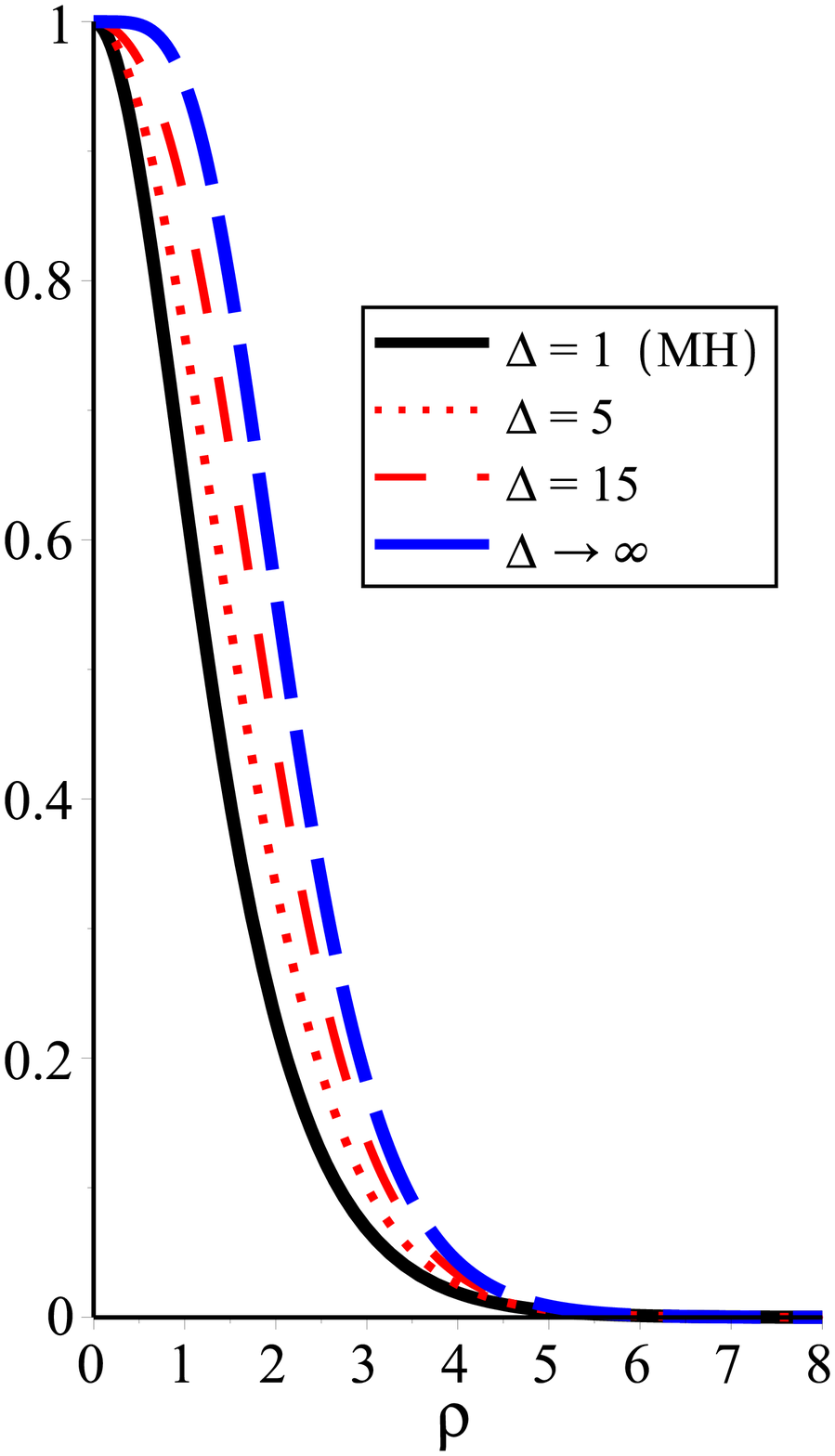}
\caption{Vector potential ${a}(\protect\rho)$.}
\label{A}
\end{figure}

Figure \ref{A} depicts the profiles of the vector potential. For $0<\Delta <1,$
the profiles are wider when $\Delta \rightarrow 0,$ and narrower when $%
\Delta \rightarrow 1$. The maximum narrowing is attained in $\Delta =1,$
where MH model (solid black line) is recover. For $\Delta >1,$ the profiles
becomes wider when $\Delta $ increases, but the incremented width has a
limit that is attained when $\Delta \rightarrow \infty $ (solid blue line).

\begin{figure}[H]
\centering
\includegraphics[width=4.35cm]{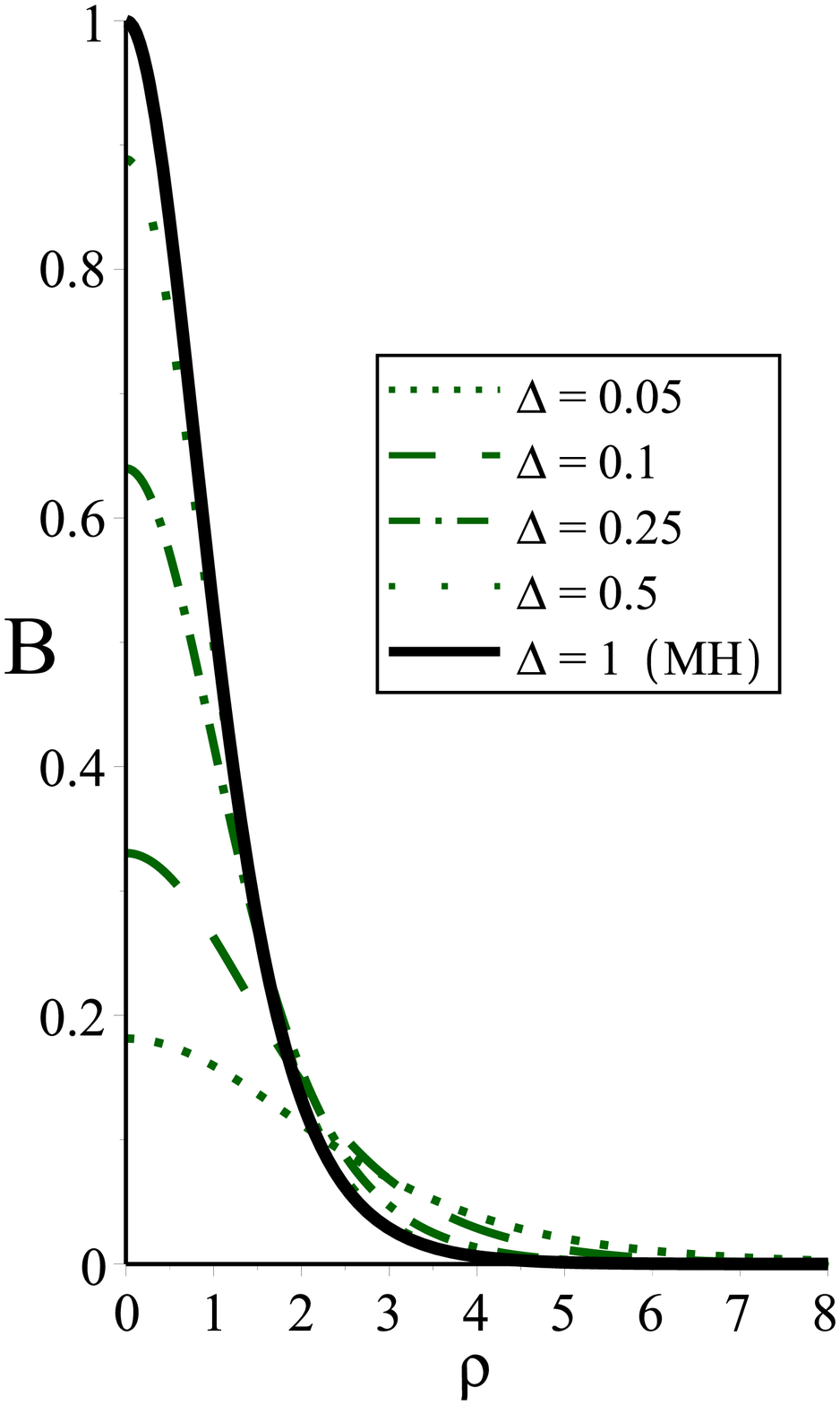}\!\!%
\includegraphics[width=4.25cm]{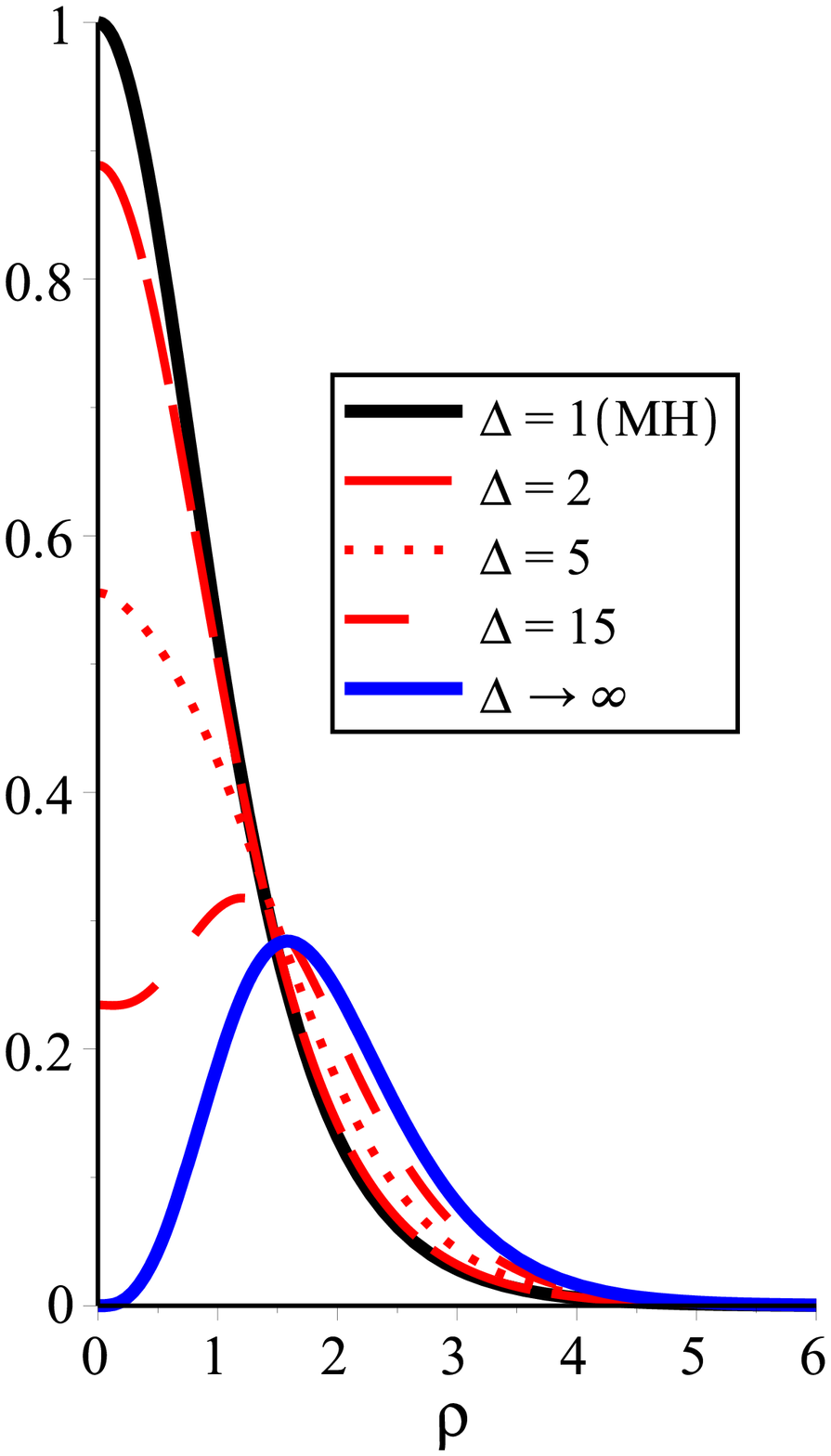}
\caption{Magnetic field ${B}(\protect\rho)$.}
\label{MF}
\end{figure}

The magnetic field profiles are shown in Fig. \ref{MF}, whose amplitudes at
origin are given by $4\Delta /(\Delta +1)^{2}$. For the range $0<\Delta \leq
1$ (green lines), they are lumps centered at origin with increasing
amplitudes whenever $\Delta $ increases, whose maximum value is reached for $%
\Delta =1$ (MH model). For $\Delta >1,$ the amplitude decreases when $\Delta
$ increases continuously. An interesting fact is observed around $\Delta
\sim 6$: the lump format of the magnetic field begins to deform, transforming
in a ringlike profile. In the limit $\Delta \rightarrow \infty ,$ it has a
null value at the origin such that the ring profiles resemble those of CSH
and MCSH models.
\begin{figure}[H]
\centering
\includegraphics[width=6cm]{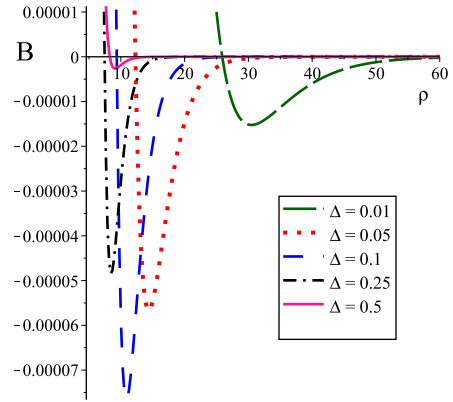}
\caption{Magnetic flux inversion.}
\label{MF INV}
\end{figure}
The model has localized magnetic flux inversion when $0\leq \Delta <1,$ such
as it is shown in Fig. \ref{MF INV}, where a zoom was performed for the
profiles with $0.01\leq \Delta \leq 0.5$. One can clearly observe the
localized magnetic flux inversion, which is more pronounced for values $%
\Delta <0.5$. For $\Delta >0.5,$ such inversion does not exist. This
peculiarity in this Lorentz-violating MH model was also observed in the absence
of LV in Higgs sector \cite{Carlison2}. Thus, we can say that such
localized magnetic flux inversion is a proper effect of CPT-even and
Lorentz-violating terms.

\begin{figure}[H]
\centering
\includegraphics[width=8.25cm]{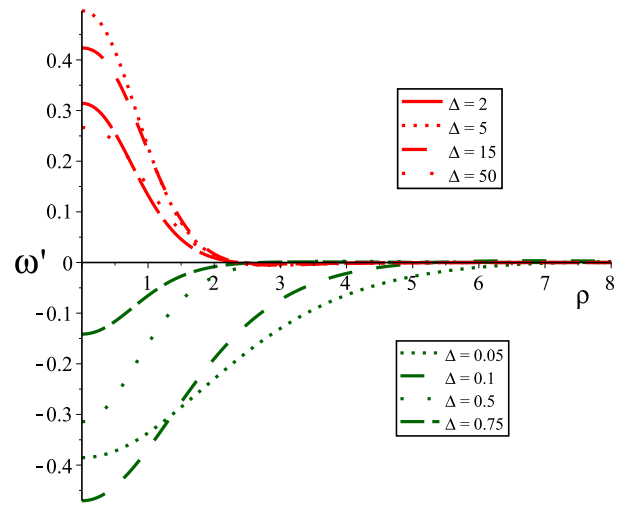}
\caption{Electric field ${\protect\omega}^{\prime}(\protect\rho)$.}
\label{EF}
\end{figure}

Figure \ref{EF} shows the electric field profiles, which are lumps centered at
origin, whose amplitudes are given by Eq. (\ref{bcsxx}). It is negative for $%
0<\Delta <1,$ taking its minimum value $-1/2,$ when $\Delta =3-2\sqrt{2}\sim
0.172,$ and it is positive for $\Delta >1,$ taking its maximum value $1/2,$
when $\Delta =3+2\sqrt{2}\sim 5.828$. The electric field is null when $%
\Delta \rightarrow 0,1,\infty $.

\begin{figure}[H]
\centering
\includegraphics[width=4.35cm]{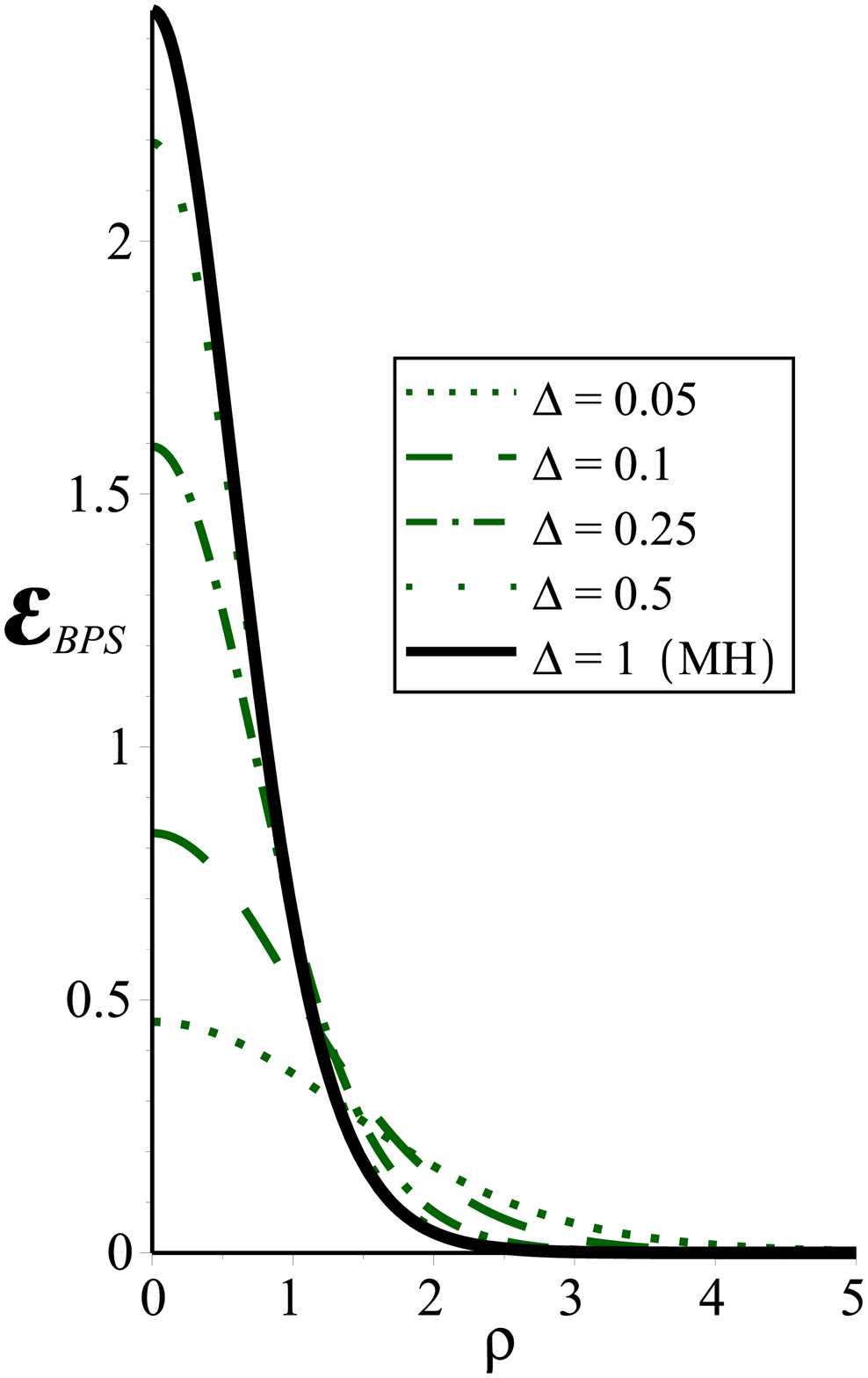}\!\!%
\includegraphics[width=4.25cm]{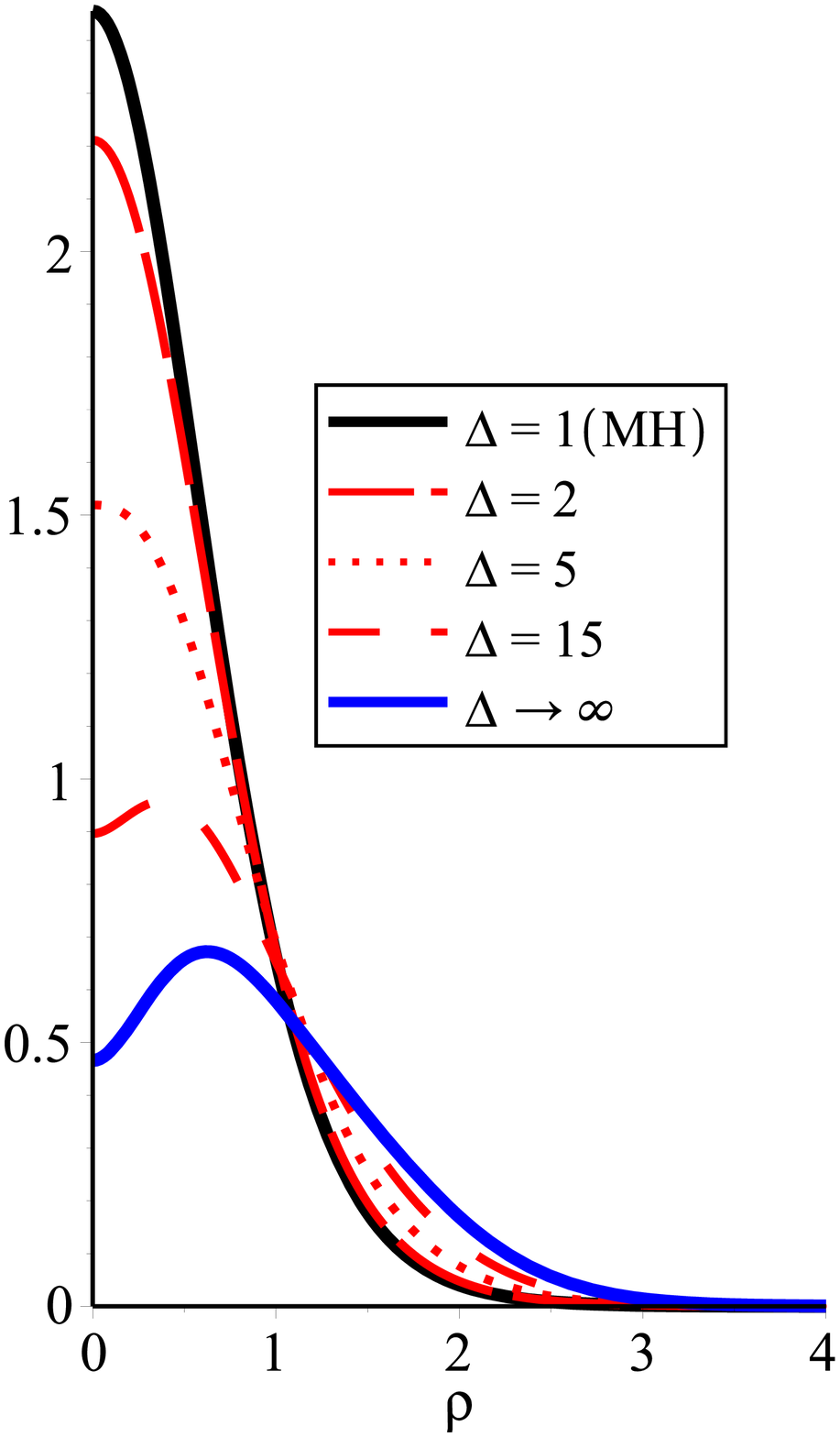}
\caption{BPS energy density ${\mathcal{E}}(\protect\rho)$.}
\label{ED}
\end{figure}

Figure \ref{ED} presents the profiles for the BPS energy density, which are
similar to magnetic field ones due to the direct relation between energy and
magnetic flux. For $0<\Delta <1$, they are lumps centered at the origin whose,
amplitude increases when $\Delta $ increases, attaining its maximum value in $%
\Delta =1$, which reproduces the BPS energy density of the MH model. For $%
\Delta >1$, the lumps decrease in their amplitudes, but around $\Delta \sim 6,$
they transform in rings. Such ring-like behavior is maintained for all
values of $\Delta >6$, and their amplitudes at origin decrease whenever $%
\Delta $ increases continuously such that in the limit $\Delta \rightarrow
\infty ,$ the amplitude is $\sim 0.46$ (but for $n>1$, the amplitude at
origin is zero). The ringlike behavior resembling the CSH and MCSH
behaviors is an interesting effect of LV in the Higgs sector.

\section{REMARKS AND CONCLUSIONS}

We have performed a comprehensive analysis on the formation of charged BPS
vortices in the context of the SME. Two cases were studied separately. First,
we have discussed the Abelian MH model supplemented with the CFJ term in
electromagnetic sector and a CPT-even Lorentz-violating term in the Higgs sector. The second
case analyzed is also the Abelian MH model but this time provided with
CPT-even and Lorentz-violating terms in both the photon and Higgs sectors.
We have verified that for supporting charged BPS vortex solutions the
original models must be modified by introducing a neutral scalar field $\Psi
$, with appropriate dynamic. This procedure is similar to the one that
happens in MCSH model. Notwithstanding, there are some differences between
the CPT-odd and CPT-even charged vortices: In the CPT-odd case, both the
charge density and the total charge are non-null whereas, in the CPT-even
case, the charge density is nonzero, but the total electric charge is null.
Another important result involves the magnetic flux, which, besides being
proportional to the winding number, also depends explicitly in the
Lorentz-violating coefficients belonging to the Higgs sector, such as
presented in Eqs. (\ref{flux_cfj}) and (\ref{flux_mhLV}). This result was
firstly obtained in the context of uncharged BPS vortices in SME \cite%
{Carlison1}, and it appears to be a general effect produced by the CPT-even
and Lorentz-violating terms of the Higgs sector.

One of the observations already made in Ref.\cite{Carlison1} is the change of
the vortex ansatz (\ref{newaz}) describing the profiles of the Higgs and
gauge fields. This is an important fact because the boundary condition
satisfied by gauge field at origin [see Eqs. (\ref{BCGD1}) and (\ref{BC})]
is modified by LV belonging to Higgs sector. Otherwise, the solutions of
self-dual or BPS equations will not be topological; neither will have finite
energy. Such change in gauge field boundary conditions are crucial for the
new characteristic presented by the quantized magnetic flux.

Our analysis shows that for fixed LV in photon sector, the LV in Higgs
sector increases the richness of BPS solutions in the context of SME. For
example, the presence of LV in Higgs sector allows to interpolate CPT-odd
vortices between very spread for $0<\Delta <1,$ to some more localized ones
when $\Delta \rightarrow 1$. For $1\leq \Delta <\infty ,$ profiles run from
the MCSH $\left( \Delta =1\right) $ to the MH ($\Delta \rightarrow \infty $)
solutions. For the CPT-even vortices, the profiles in the region $0<\Delta
\leq 1,~$are wider and more spread for decreasing $\Delta $. But now the
value $\Delta =1,$ represents the vortex solution of MH model. For $\Delta
>1$ and increasing $\Delta $, the vortex profiles are confined between the MH
model (solid black line) and the one obtained when ($\Delta \rightarrow
\infty $) (solid blue line). An interesting fact in CPT-even vortices is the
appearance of ringlike profiles (for magnetic field and BPS energy density)
for sufficiently large values of $\Delta $ parameter, as explicitly shown in
Figs. \ref{MF} and \ref{ED}, respectively. Another peculiarity is the
localized magnetic flux inversion, such as was also observed in the
absence of LV in Higgs sector \cite{Carlison2}. This property appears to
be a typical effect of CPT-even and Lorentz-violating terms in photon sector.

A remark on the subject of considering large values for
Lorentz-violation coefficients: In the context of standard model extension, the Lorentz-violating parameters
must be sufficiently small. However, as energy positivity is preserved
whenever $\Delta >0$, we can consider large values of $\Delta $ within an
effective model describing the electrodynamics of some type of material
media.

In this context we, can also comment about implication of
these Lorentz-violating models in superconductivity physics. The new {%
materials presenting high-temperature superconductivity are
anisotropic; most of them have a layered structure. The
simplest GL model is for an uniaxial superconductor whose free energy
density is written as%
\begin{eqnarray}
\mathcal{F} &=&\frac{B^{2}}{8\pi }+\alpha \left\vert \phi \right\vert ^{2}+%
\frac{\beta }{2}\left\vert \phi \right\vert ^{2}  \label{ffee} \\
&&\hspace{-0.5cm}+\gamma _{jk}\left( i\hbar \partial _{j}\phi ^{\ast }-\frac{%
2e}{c}A_{j}\phi ^{\ast }\right) \left( -i\hbar \partial _{k}\phi -\frac{2e}{c%
}A_{k}\phi \right) ,  \notag
\end{eqnarray}%
where $B$ is the magnetic field produced by the gauge
field }$A_{k}$ and $\beta $ is a constant, but $\alpha $%
 depends in temperature changing its sign at $T=T_{c}$,
and $\gamma _{jk}$ is a constant tensor coupled to the crystal
axes.

It is possible to establish a connection between the
Lorentz-violating models whose energy densities are given by Eqs. (\ref%
{Hstat}), (\ref{E_X}) and the free energy (\ref{ffee}) describing a high-$
T_{c}$ superconductor. Such connection is provided by the term  $%
\left[ \delta _{ij}-\left( k_{\phi \phi }\right) _{ij}\right] \left(
D_{i}\phi \right) ^{\ast }\left( D_{j}\phi \right) $, which allows
to relate the matrix  $\delta _{ij}-\left( k_{\phi \phi }\right) _{ij},$%
 with the tensor $\gamma _{jk}.$ It is clear that in both
LV models, the coefficients do not depend in temperature as
happens in phenomenological Ginzburg-Landau's models. The study of possible
contributions to superconductivity can be obtained by analyzing the
nonrelativistic limits of both LV models at finite temperature. Such
approach would allow to analyze Lorentz-violations effects or contributions
in London free energy, the Josephson effect, the interaction of vortices with
the crystalline structure (the vortex's intrinsic pinning), and resistivity, among
others. Advances in such interesting research would be developed in a future
manuscript.

Future developments can be oriented to analyze the influence of
Lorentz violation in the existence of topological defects in the non-Abelian
sector of the standard model-- i.e., magnetic monopoles and non-Abelian
vortices in the context of standard model extension. The results will be
reported soon.

\textbf{ACKNOWLEDGMENTS}

The authors thank CAPES, CNPq and FAPEMA (Brazilian agencies) for
financial support.

\end{document}